\documentclass[twocolumn]{aastex63}  
\usepackage{color}

\accepted{March 5, 2021}

\submitjournal{ApJS}
\shorttitle{How to Train Your Flare Forecasting Models}
\shortauthors{Ahmadzadeh et al.}

\begin{document}

\title{How to Train Your Flare Prediction Model:\\Revisiting Robust Sampling of Rare Events}

\correspondingauthor{Azim Ahmadzadeh}
\email{aahmadzadeh1@gsu.edu}

\author[0000-0002-1631-5336]{Azim Ahmadzadeh}
\affiliation{Georgia State University \\
Atlanta, GA 30302, USA}

\author[0000-0002-9799-9265]{Berkay Aydin}
\affiliation{Georgia State University \\
Atlanta, GA 30302, USA}

\author[0000-0001-6913-1330]{Manolis K. Georgoulis}
\affiliation{RCAAM of the Academy of Athens, \\
Soranou Efesiou 4, Athens, GR-11527, Greece}

\author[0000-0002-1837-8365]{Dustin J. Kempton}
\affiliation{Georgia State University \\
Atlanta, GA 30302, USA}

\author[0000-0003-1753-8002]{Sushant S. Mahajan}
\affiliation{Institute for Astronomy, University of Hawai`i at M\={a}noa, \\
Honolulu, Hawaii, USA}

\author[0000-0001-9598-8207]{Rafal A. Angryk}
\affiliation{Georgia State University \\
Atlanta, GA 30302, USA}

\begin{abstract}
    We present a case study of solar flare forecasting by means of metadata feature time series, by treating it as a prominent class-imbalance and temporally coherent problem. Taking full advantage of pre-flare time series in solar active regions is made possible via the Space Weather Analytics for Solar Flares (SWAN-SF) benchmark dataset;  a partitioned collection of multivariate time series of active region properties comprising 4075 regions and spanning over 9 years of the Solar Dynamics Observatory (SDO) period of operations. We showcase the general concept of temporal coherence triggered by the demand of continuity in time series forecasting and show that lack of proper understanding of this effect may spuriously enhance models' performance. We further address another well-known challenge in rare event prediction, namely, the class-imbalance issue. The SWAN-SF is an appropriate dataset for this, with a 60:1 imbalance ratio for GOES M- and X-class flares and a 800:1 for X-class flares against flare-quiet instances. We revisit the main remedies for these challenges and present several experiments to illustrate the exact impact that each of these remedies may have on performance. Moreover, we acknowledge that some basic data manipulation tasks such as data normalization and cross validation may also impact the performance – we discuss these problems as well. In this framework we also review the primary advantages and disadvantages of using true skill statistic and Heidke skill score, as two widely used performance verification metrics for the flare forecasting task. In conclusion, we show and advocate for the benefits of time series vs. point-in-time forecasting, provided that the above challenges are measurably and quantitatively addressed.

\end{abstract}


\section{Introduction} \label{sec:intro}

    Data collected for addressing real-world problems is almost never clean and ready to use, no matter how carefully a screening process was carried out. Inevitably, there are some challenges inherited either by the nature of the studied subject or by the data collection strategy, that should be identified, understood, and dealt with accordingly. Class imbalance and temporal coherence are two of such challenges that are present in many nonlinear dynamical systems such as earthquake prediction, fraud detection, or weather forecasting. This study revisits their impact in another natural manifestation of these issues, namely solar flare prediction.
    
    Solar flares are sudden and substantial enhancements of radiation spanning over the entire electromagnetic spectrum, including its high-energy part (extreme ultraviolet, X-rays, $\gamma$-rays). They occur at local scales in the Sun and pose a threat to humans and equipment in space. Since 1974, X-ray flares are automatically detected and classified by the National Oceanic and Atmospheric Administration's (NOAA) GOES satellites in the 1 - 8 Angstrom wavelength range. Based on peak soft X-ray flux in this range, flares are logarithmically classified as A, B, C, M, and X, from weaker to stronger, starting from $10^{-8} W/m^2$. Therefore, an X-class flare is generally ten times stronger, in terms of peak flux, than an M-class flare and 100 times stronger than a C-class flare. Within each class there is a subclass denomination from 1 to 9 \citep[see, for example,][and resources therein, for a tutorial]{fox2011flares}. When the X-ray background level is high, as typically occurs at times of elevated solar activity, A- and B-class flares are often difficult or impossible to discern while C-class flares and above are mostly detected, particularly above level C2 \citep{wagner_88,veronig_etal04,aschwanden_freeland12}. The most intense flares, namely the M and X classes, are commonly targeted for prediction due to their potentially adverse space-weather ramifications. In spite of more than 20 years of research and meaningful advances, flare prediction remains a largely outstanding problem \citep[e.g.,][]{LEKA201865}.

    Some of the major challenges the flare forecasting researchers are up against are rooted in the rarity of the events of interest, the high-dimensionality of observational data, and the dynamic behavior of the Sun. Below we briefly review these challenges and a number of relevant studies.
    
    \noindent\textbf{Extreme Class Imbalance.}
        The frequency distribution of the peak X-ray fluxes of solar flares is a robust power law with a dynamical range spanning several orders of magnitude. A common interpretation of flares views them as stochastic manifestations of self-organized criticality \citep[see, for example, the comprehensive review of][and numerous references therein]{aschwanden2016Instabilities}.
            
        A statistical analysis of the flares reported by the National Oceanic and Atmospheric Administration (NOAA) during solar cycle 23 (1996 to 2008) shows that around $50\%$ of active regions produce at least one C-class flare, with approximately $10\%$ of them producing at least one M-class flare and less than $2\%$ producing at least one X-class flare \citep{georgoulis2012b}. Solar cycle 24 (2009 to present), according to the dataset created by \citet{angryk2020multivariate}, exhibits a weaker major flare crop at a similar number of active regions to solar cycle 23 further highlighting the importance of effectively treating  class imbalance. 

    \noindent\textbf{Point-in-time vs. Time Series Forecasting.}
        Solar flare forecasting has been humanity's first attempt toward space weather forecasting. As such, numerous magnetic properties and forecast methods have been proposed since the early 1990's 
        \citep[see, e.g.,][for non-exhaustive reviews]{georgoulis2012our,Barnes_2016}
        while studies facilitating several predictive properties were performed by many others \citep[see, for example, ][]{2003ApJ...595.1296L,2007ApJ...656.1173L,qahwaji2007automatic,barnes2008evaluating, welsch2009what,georgoulis2012b,bobra2015solar}. The first comprehensive comparison between multiple flare prediction methods was undertaken by \citet{2016ApJ...829...89B}, while comparative evaluations between operational and pre-operational methods were implemented by \citet{2019ApJS..243...36L,2019ApJ...881..101L,park_etal2020}.  
            
        A quick perusal of the voluminous literature, however, will show that the majority of these methods correspond to point-in-time forecasting, namely, to using the instantaneous value of one or more parameters in order to produce a binary or probabilistic flare forecast over a preset forecast horizon. But flares are an inherently dynamical phenomenon, with clear pre-flare and post-flare phases, characterized by certain evolutionary trends \citep[e.g., ][]{benz2008flare, fletcher2011observational}. Because of this, time series of candidate flare forecast parameters should be used, rather than isolated points in time. Quite likely, the level of difficulty in assimilating time series of predictive parameters in flare forecasting was the main reason for utilizing (possibly over-) simplified point-in-time forecasts in many cases. This compromise, however, may have hampered further progress: early works as well as recent ranking efforts of predictive flare parameters show that previous flare history is a significant factor to consider \citep{2003ApJ...595.1296L, barnes2008evaluating,falconer2012prior, 2019ApJS..243...36L, features2019Campi,park_etal2020}, hinting toward the need to study the pre-flare temporal evolution in terms of both parameter time series and flare history.
            
    \noindent\textbf{Non-representative Datasets.}
        While there is no shortage of ground- and space-based telescopes mapping the magnetic field of the Sun's photosphere over the past 30 years \citep{cacciani1990observations, bala_west91,scherrer1995solar, lites96, mickey_etal96, spirock2001big, tsuneta2008solar, scherrer2012helioseismic} and new instruments are just about to be commissioned \citep{rimmele_etal20, solanki_etal20} it is clear that (i) the temporal span of space-based vector magnetographic data is still limited to one solar cycle and (ii) training forecast methods on certain parts of a solar cycle is not necessarily optimal for forecasting other parts of the same and/or different cycles, due to the continuously modulating background of magnetic activity \citep[see, for example, the discussion on varying flare occurrence, or climatology, in ][]{2018JSWSC...8A..34M,2019ApJS..243...36L}. This heterogeneity, coupled with random undersampling for majority class events, causes the sampled subsets of data to become non-representative of the overall flare population. Therefore, in many cases, the performance of forecasting models may critically depend on the sampling strategy.
            
        Each of the challenges mentioned above requires a proper treatment. Some correspond to the data collection phase, some to the cleaning and preprocessing phase, and some to the learning phase. While many studies have revisited these challenges \citep[to name a few in geo- and space science domains, see ][]{woodcock1976evaluation,Bloomfield_2012,camporeale2019challenges}, there is an obvious  emphasis by interested communities on the performance of forecast models. Less obvious, however, is that this emphasis often comes at the expense of the models' robustness. Typical forecast metrics used for performance verification in flare forecasting models are the true skill statistic (TSS) and realizations of the Heidke skill score (HSS; HSS2) - see Section \ref{sec:verificationMeasures}. Recent works with apparently high, but perhaps not particularly robust, performance include \citet{hamdi2017timeseries} (TSS$\geqslant$0.8 or even $\geqslant$0.9 for a 24-hour prediction of flares greater than M1.0) or \citet{Nishizuka_2017} (TSS$\geqslant$0.9 for similar forecast settings). In other well-cited studies \citep[for a comparison, see the reports in][]{bobra2015solar,Barnes_2016}, the highest TSS that maintains a similar HSS2 varies around $\approx$0.6. We, therefore, understand that without a major change in the learning process, exceedingly high TSS values may simply be a result of sub-optimal preprocessing. Some factors crucial to a model's performance verification are preprocessing steps, sampling strategies, and the choice of performance evaluation metrics. A model's performance varies when any of these factors undergoes changes. However, a robust model is impacted only minimally relative to others as a result of these changes. That is, a robust model is less prone to overfitting; a known phenomenon that describes a model that is too closely fit to a limited set of data points and therefore does not generalize well. Throughout this study, we frequently draw a comparison between robustness and performance regarding different practices.
            
        The rest of this paper is organized as follows: In Section \ref{sec:theData}, we first briefly introduce the SWAN-SF; the dataset we run our experiments on. This dataset allows us to showcase the concept of the class imbalance (Section \ref{subsec:classImbalance}) and temporal coherence (Section \ref{subsec:temporalCoherence}) of data. We also discuss data normalization (Section \ref{subsec:dataNormalization}) and hyperparameter tuning (Section \ref{subsec:hyperparameterTuning}) of models. Section \ref{sec:forecastDataset} then discusses the choice of the forecast dataset from the SWAN-SF, while in Section \ref{sec:modelAndTreatments} we present some of the major treatments regarding class-imbalance (Section \ref{subsec:classImbalanceTreatments}) and temporal-coherence (Section \ref{subsec:temporalCoherenceTreatments}) issues. Before presenting our experiments and the results, we briefly review in Section \ref{sec:verificationMeasures} the performance verification metrics we use. In Section \ref{sec:experiments}, we discuss the details and results of a number of experiments we designed for examining the above-mentioned challenges and illustrate the impact of the corresponding treatments. We conclude in Section \ref{sec:summary} by listing the key lessons learned from these experiments.

\section{Benchmark Dataset and Challenges}\label{sec:theData}
    In this section, we briefly discuss the SWAN-SF benchmark dataset of \citet{angryk2020multivariate} and the major issues intrinsic to this and similar datasets. If not properly treated, these issues may well lead to overestimation of the forecasts' performance.
    
    \subsection{SWAN-SF Benchmark Dataset}\label{subsec:swan-sf}
        Multiple flare prediction studies \citep{Barnes_2016,bobra2015solar,Bloomfield_2012} and the European Union FLARECAST project \citep{benvenuto2018hybrid,florios2018forecasting,features2019Campi} emphasize machine learning for flare prediction, but they use point-in-time measurements. Given different periods of observation and partitioning/sampling strategies, we are unable to determine whether the potential differences in accuracy or skill score values in these studies reflect the intrinsic stochasticity in flare occurrence, the specific preprocessing and sampling strategies carried out, the implementation of machine learning models, or perhaps a combination of the above. This is the core reason behind the development of the Space Weather Analytics for Solar Flares (SWAN-SF) benchmark dataset, composed entirely of multivariate time series. Its purpose is to facilitate unbiased flare forecasting and hopefully make strides toward non-incremental future improvements in forecasting.
        
\begin{figure}[t]
            \centering\includegraphics[width=\linewidth]{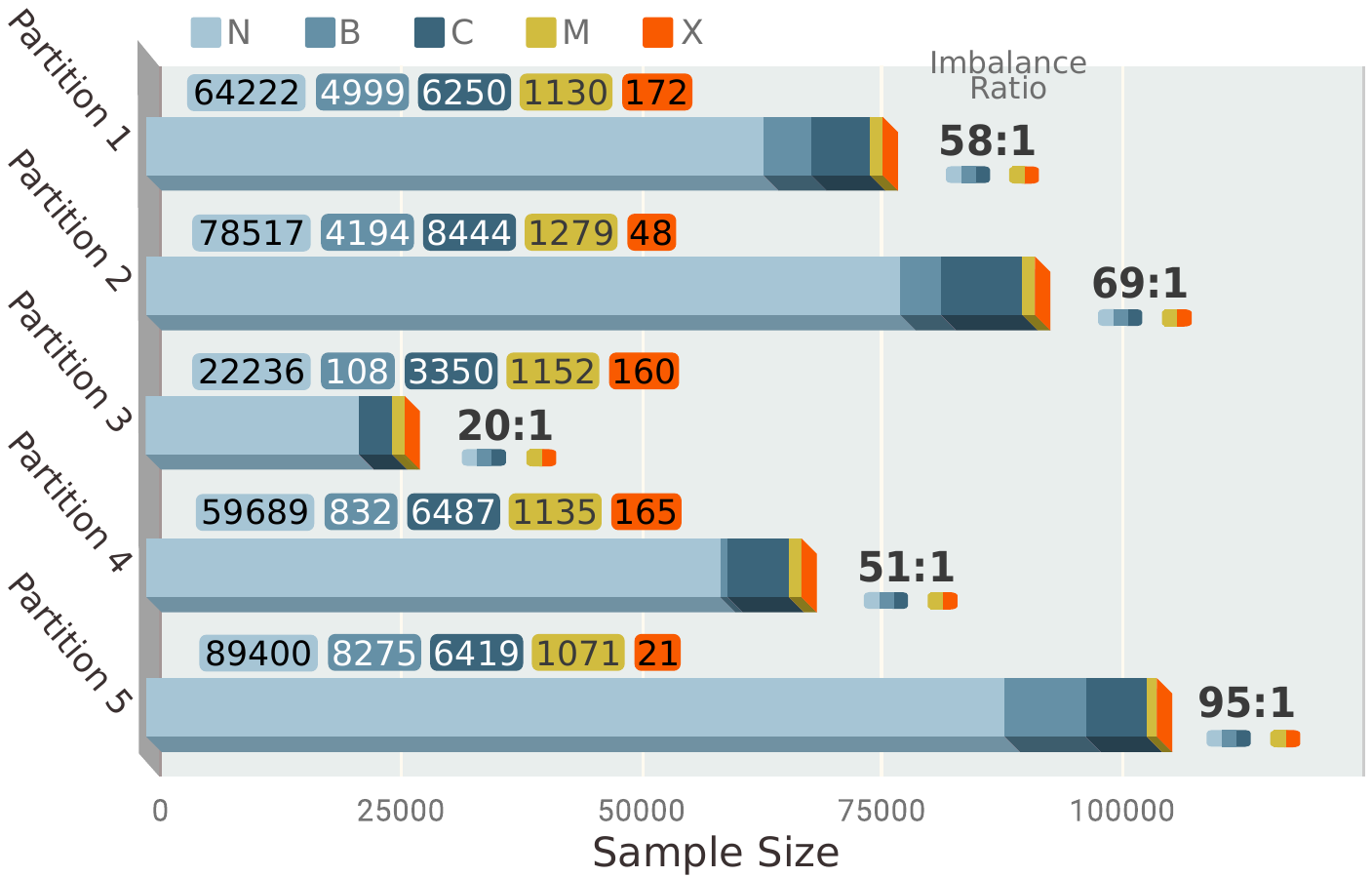}
            \caption{Stacked bar plot of multivariate time series populations for each of the five flare classes across different partitions of the SWAN-SF benchmark dataset. Flare classes X, M, C and B are taken and validated from GOES classification, while N denotes flare-quiet and GOES A-class instances. The annotated populations are stemming from the current time series slicing strategy of the SWAN-SF with a 1-hour step size, a 12-hour observation window, and a 24-hour prediction window. The sliced multivariate time series are labeled with the class of the strongest reported flare, if any, within the respective prediction window.}
            \label{fig:classfrequencies}
        \end{figure}

        The data points of the SWAN-SF benchmark dataset are labeled by $5$ different flare classes, namely GOES X, M, C, B, and a non-flaring class denoted by N. Therefore, N includes flare-quite instances and GOES A-class events. We hereafter omit GOES from the above classes, for brevity. This dataset comprises five temporally segmented partitions and is designed in such a way that each partition includes approximately the same number of X- and M-class flares (Fig.~\ref{fig:classfrequencies}). The dataset contains various time series parameters derived from solar photospheric magnetograms along with NOAA's flare history of active regions. Magnetograms and their metadata are provided by the Solar Dynamics Observatory's \citep{Pesnell2012} HMI Active Region Patches (HARP) data product \citep{hoeksema2014harp}. The magnetic field parameters are physics-based and were originally taken by the Space weather HMI Active Region Patches (SHARP) data product \citep{bobra2014helioseismic}), but were recalculated for validation purposes and enhanced with parameters not present in SHARPs (see Table 1 in \cite{angryk2020multivariate}).
        
        The data points in this dataset are time series slices of parameters extracted from solar active regions in a sliding fashion. That is, for a particular flare with a unique id, $k$ equal-length multivariate time series are collected from a fixed period of time in the pre-flare history. This is called the \textit{observation window}, $T_{obs}$. Denoting $s_i$ as the starting point of the $i$-th slice of the multivariate time series, the $(i+1)$-th slice starts at $s_i + \tau$, where $\tau$ is the step-size in the sliding process. Each of the extracted multivariate time series is assigned a label, determined by the class of the strongest flare recorded during a fixed-size temporal window that follows $T_{obs}$. This window is the \textit{prediction window}, $T_{pred}$, and starts precisely at the end of $T_{obs}$. In the SWAN-SF, $T_{obs}$ and $T_{pred}$ span over $12$ and $24$ hours, respectively, with $\tau = 1$ hour. Placing the prediction window right after the observation window implies a zero latency (i.e., a finite time before a given forecast comes into effect) in our tests, for simplicity.
    
    \subsection{Class Imbalance of Data}\label{subsec:classImbalance}
        We characterize a dataset as class imbalanced when the population of one or more data classes are significantly smaller than those of the majority classes. The data points of the smaller group are called the \textit{minority} instances. If forecasting is intended for them, they are also called \textit{positive} instances. Conversely, the other group's data points are called the \textit{majority}, or \textit{negative}, instances, in case minority instances are the forecast targets. The SWAN-SF benchmark dataset exhibits extreme class-imbalance: the imbalance ratios for each partition of the dataset and for using GOES M- and X-class flares as the  minority class are annotated in Fig.~\ref{fig:classfrequencies}.

        Class-imbalance is a well-known concept in the Machine Learning and Statistics communities  \citep{kubat1997addressing, japkowicz2002class, ganganwar2012overview, krawczyk2016learning} but has been also addressed in many other domains, including atmospheric sciences and flare forecasting \citep{woodcock1976evaluation,jolliffe2012forecast,Bloomfield_2012,bobra2015solar}. However, the complexity of the forecasting problem sometimes causes this important issue to go relatively unnoticed resulting in a large variance between the reported performance of the models, as we briefly pointed out in the Introduction. 
        
        \begin{figure*}[t]
            \centering\includegraphics[width=0.8\textwidth]{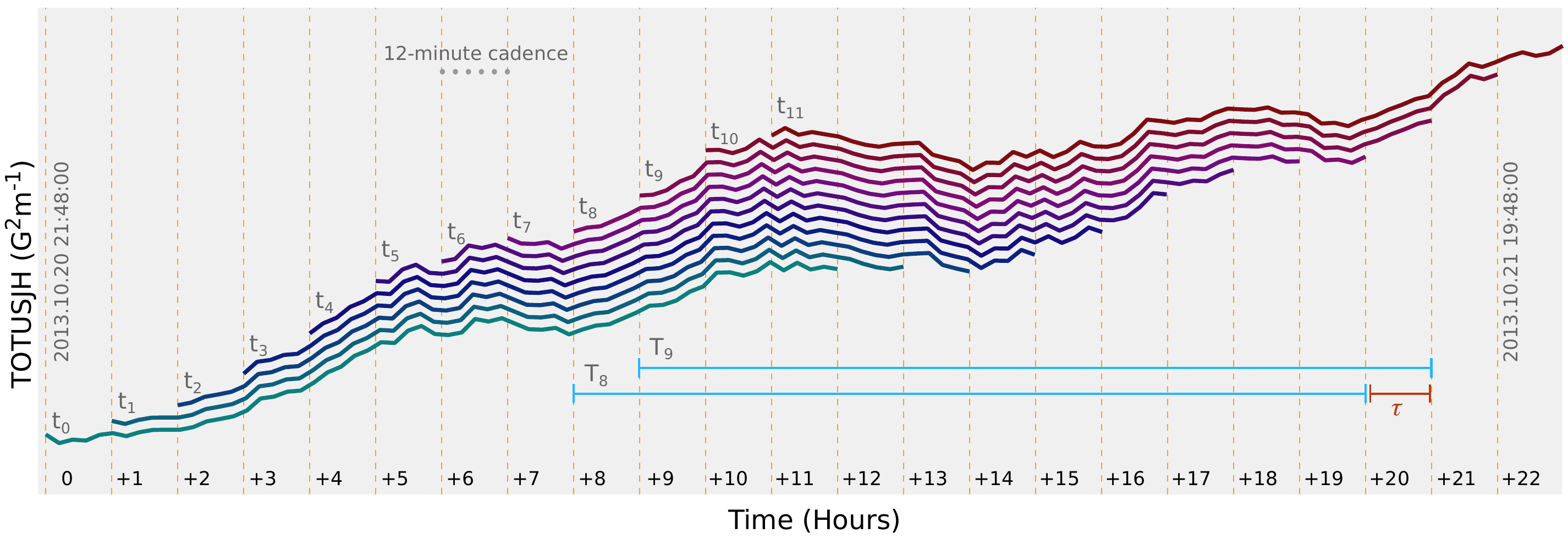}
            \caption{Twelve consecutive time series slices for the parameter \textit{Total Unsigned Current Helicity} (TOTUSJH) corresponding to an M1.0-class flare associated to NOAA AR 11875 (HARP 3291). Each time series spans over 12 hours of observation, with a 12-minute cadence. The blue interval segments $T_8$ and $T_9$, for example, show the domain of time series $t_8$ and $t_9$, respectively, with  $\tau$ representing the slicing step-size, i.e., 1 hour. The vertical spacing of each time series slice along the $y$-axis has been added manually to make different time series discernible.}
            \label{fig:temporalCoherence}
        \end{figure*}

        The class-imbalance issue can be roughly explained as follows: classification models, in general, aim to optimize their cost functions by minimizing the total number of misclassifications. Given the significantly higher density of the majority class, a correct classification on the decision boundary, where two or more classes overlap, is typically accompanied by multiple misclassified majority instances. The Support Vector Machine (SVM) classifier \citep{vapnik1963pattern}, for example, calculates an optimal hyper-plane that makes this separation \citep{burges1998tutorial,ben2010user}. Due to the imbalance between the classes, a hyper-plane that is (ideally) supposed to pass through the decision boundary will be shifted further into the region of the minority class (classifying almost all the data points as the majority class) to reduce the total number of incorrect classifications. This results in higher true-negatives (i.e., correct classification of CBN-class flares) and lower true-positives (i.e., correct classification of XM-class flares as per the example of Fig.~\ref{fig:classfrequencies}). In other words, a model in a class-imbalanced space always tends to favor the majority class. This is of particular concern because flare forecasting research focuses on the minority, rather than majority, instances.
    
        Another angle to this problem is the determination of appropriate performance measures. Many well-known performance metrics are significantly impacted by class imbalance \citep{hossin2015review}, including \textit{accuracy}, \textit{precision} (but not \textit{recall}), and hence the \textit{F1-score}. This is mainly because these measures ignore the number of misclassifications. For instance, a na\"ive model that classifies all instances as the negative (majority) class may result in a very high (often asymptotic to $1.0$) accuracy, while learning very little or nothing about the minority class. In Section \ref{sec:verificationMeasures}, we review some popular metrics used in flare forecasting. Furthermore, in Section \ref{subsec:classImbalanceTreatments} we discuss the classical approaches to tackle the class imbalance issue.

    \subsection{Temporal Coherence of Data}\label{subsec:temporalCoherence}
        The sliding observation window mentioned in Section \ref{subsec:swan-sf} is crucial for operational prediction on real-time data. Indeed, a continuous and uninterrupted observation is needed for training a forecast model. This continuity dictates the existence of partially overlapping slices of multivariate time series, and therefore, non-independent data points. A characteristic example is provided in Fig.~\ref{fig:temporalCoherence}, where 12 consecutive 12-hour slices of the parameter \textit{Total Unsigned Current Helicity} (\textit{TOTUSJH}), extracted from an active region with HARP number $3291$, are visualized. From this example, one clearly notices that a physical parameter of an active region may not be expected to behave significantly differently from a slice to its adjacent one. Any two adjacent time series are identical for a period of $11$ hours, and that accounts for more than $90\%$ (precisely, $\frac{11}{12}=0.92$) of each time series' length. Of course, this is not specific to the particular parameter used in the example, but is a characteristic inherited from the one-hour time-stepped slicing methodology.

\begin{figure}[t]
            \centering\includegraphics[width=1.0\linewidth]{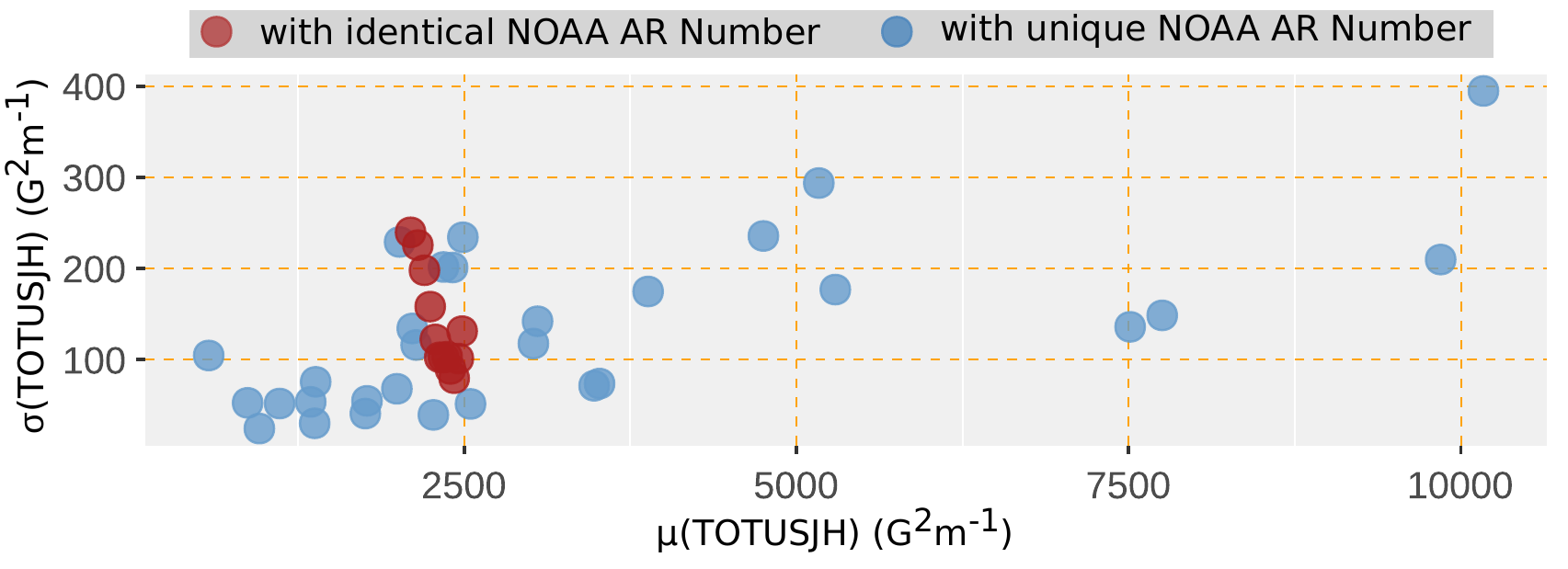}
            \caption{Standard deviation ($\sigma$) vs. mean ($\mu$) of parameter TOTUSJH, for two different groups: one (red/darker circles) extracted from the consecutive slices of Fig.~\ref{fig:temporalCoherence} and another (blue/lighter circles) from a randomly selected set of time series slices for different active regions. Both groups correspond to M1.0-class flare labels and stem from Partition 3 of the SWAN-SF. The apparent clustering of red circles is due to the temporal coherence of consecutive time series.} 
            \label{fig:temporalCoherenceExtracted}
        \end{figure}

        Let us now look at the example of Fig.~\ref{fig:temporalCoherence} more closely: these very similar slices, identical for 11 hours, will appear as a cluster of data points in the multi-dimensional feature space, located close to each other. This clustering is apparent in Fig.~\ref{fig:temporalCoherenceExtracted}, where we plot the \textit{mean} vs. the \textit{standard deviation} of the TOTUSJH slices of Fig.~\ref{fig:temporalCoherence} for the same active region (red/darker circles), and compare them with assorted means and standard deviations from different active regions on the same parameter (blue/lighter circles). For an adequate comparison, in terms of flare label and solar cycle phase, all blue instances correspond to the same flare label (M1.0) and partition (partition 3). As expected, the \textit{mean} values of the time series sampled from the same active region (red/darker circles) are confined to a very narrow subspace, i.e., limited to the interval $(2300, 2400) (~G^{2}m^{-1})$ of the \textit{mean} dimension. Any forecast method treating this as a potentially useful pattern in the data is erring in this case, because this `pattern' is simply a byproduct of the slicing strategy used to create the benchmark data. Classification algorithms are
        generally designed to decipher such issues. To avoid misleading them by introducing artificial clusters, nonetheless, data points from the same active region (such as the red points of Fig.~\ref{fig:temporalCoherenceExtracted}) should not be considered as entirely distinct and independent. 
        
        The above discussion, as the reader may have already noticed, can be traced back to the well-known assumption that random variables must be \textit{independent and identically distributed} (i.i.d) for the models to make sense. Although this practice may not look like a challenge, it is easy to be overlooked on multi-class, multivariate time series data. Randomly sampling and shuffling the instances is often assumed to tackle the problem but, as our example showed, this may not be always the case. More importantly, when such clusters are allowed to exist in both the training and validation sets, the high performance on the \textit{unseen} data (of the validation set) may convince researchers that the classifier was able to generalize the learned patterns and forecast flares. Such a superficial performance is exemplified in Experiment D of Section \ref{sec:experiments}. It is also important to note that as the dimensionality of the feature space increases (by including more statistical features), spotting such clusters becomes more and more difficult; in high dimensional data, it is very likely that many dimensions are irrelevant but they mask the existing clusters \citep{parsons2004subspace}.

        While this concept is generally known as the independence of random variables, we prefer to use our term, \textit{temporal coherence}, for two reasons: first, while the typical solution for preserving the independence between data points is random sampling, this action does not resolve the issue in this special case. Second, we would like to avoid the generality of the term `independence’ and instead point to the direct cause of the issue, i.e., the temporal overlap. We, therefore, refer to a dataset as \textit{temporally coherent}\footnote{The concept is here introduced in the context of data manipulation, not to be confused with `temporal coherence' met in Optics or elsewhere.} if (a) it comprises partially identical samples and (b) this overlap is caused by the slicing methodology used to create the dataset, regardless of it's nature, or the topic of study. 
        
        Looking at the consecutive slices shown in Fig.~\ref{fig:temporalCoherence}, it is evident that the two conditions of temporal coherence are satisfied in the SWAN-SF because: (a) the temporally-close time series are partially identical, i.e., $t_i$ and $t_{i+1}$ are identical within an 11-hour interval out of the 12 hours they each span over, and (b) this is caused by the slicing of parameter time series designed to ensure both continuity and detailed coverage, by allowing a time step much smaller than the observational window (i.e., 1 hour vs. 12 hours). Therefore, the SWAN-SF can be described as a temporally coherent dataset and the independence of data points should not be presumed. Note that we use the term `partially identical' (instead of `similar') to emphasize that what defines the temporal coherence is the temporal overlap and not the correlation between time series' values. The former is often a byproduct of the slicing methodology in the data preparation process whereas `similarity' generally refers to the homogeneity among observations.
        
    \subsection{Data Normalization}\label{subsec:dataNormalization}
        The ranges of all feature vectors of a dataset are typically transformed into a unified range before the training phase starts. This preprocessing step is called normalization and is mainly performed because it allows models to distinguish between different patterns and structures independent of the physical units and dynamical ranges of each parameter. The simplicity of this concept often leads to overlooking the variable impact of different ways of normalization \citep{shalabi2006normalization, yu2009research}. Regardless of which transformation function one may use (e.g., linear, nonlinear, or data-driven), normalization can be done `locally' or `globally'. A global normalization takes into account the global statistics, i.e., $min$ and $max$ of each feature/parameter over the entire dataset, whereas a local transformation may consider local extrema of different subsets (e.g., partitions, or training and testing sets) to normalize each of those subsets separately. Although it is common practice to apply global normalization, for the SWAN-SF and similar data sets this should be put to test due to the modulating event occurrence rate throughout solar cycle 24. In such cases, a global normalization could impact some features negatively by transforming their values into a too narrow or too wide range. We illustrate this in Experiment E in Section \ref{subsec:normalizationImpact}, by showing the significantly different performance levels using different normalization strategies.

    \subsection{Hyperparameter Tuning}\label{subsec:hyperparameterTuning}
        Temporal coherence affects also the way we tune the hyperparameters of our models. Any supervised learning model requires optimization of its hyperparameters in a data-driven manner. Since the hyperparameters should remain optimal over the entire dataset (including the incoming data points, for an operational model) the training and validation sets must each be representative of the entire dataset. In our dataset, because of the temporal coherence, random sampling does not produce such subsets (See Experiment D of Section \ref{subsec:temporalCoherenceImpact} that examines the impact of temporal coherence). Hence, the tuning process remains confined to the partitions. In Section \ref{subsec:temporalCoherenceTreatments}, while discussing a remedy for temporal coherence, we review the use of NOAA AR Number that addresses this issue. However, it is still very likely that a model highly optimized on one partition corresponding to a particular phase of a solar cycle, performs better than one that is optimized globally \citep{Barnes_2016, 2019ApJ...881..101L}. We believe hyperparameter tuning is a problem yet to be thoroughly investigated since at this point it is rather clear to us that flare forecasting has a dynamic behavior, for which ensemble models, that can be trained differently for different periods of a solar cycle, may be more appropriate \citep[for example, see][]{2015SpWea..13..626G, murray2018importance,2020JSWSC..10...38G}.

\section{Derived Forecast Dataset}\label{sec:forecastDataset}
    To use the SWAN-SF benchmark dataset, two general approaches might be taken. One is to preprocess the time series and feed them directly to supervised models for prediction. The other is to extract a set of statistical features describing the time series and then train the supervised models on these descriptors. We choose the second approach in order to carry out our experiments without having to deal with the high dimensionality of the original \added{data type, i.e.,} time series\deleted{ or the complexity of a detailed feature selection procedure}.

    \subsection{Feature Extraction}
        SWAN-SF is designed to rely entirely on time series forecasting of solar flares, if so desired. In a wealth of studies on flare forecast, only a relative minority of them have attempted to use time series \citep[e.g.,][]{2003ApJ...595.1277L, 2003ApJ...595.1296L, 2006ApJ...646.1303B, welsch2009what, reinard_etal10, cinto_etal20}. More recently, several studies moved toward time series forecasting \citep{hostetter2019understanding, ahmadzadeh2019Rare, ma2020evaluation, ji2020all, chen2020effectiveness}. In spite of a number of efforts, reports on which time series parameters perform better than others are not necessarily compelling. Since it is not the focus of this work to determine which time series characteristics will perform well, without loss of generality and in the interest of simplicity, we choose a limited set of four descriptive statistics of the time series for our experiments, namely, \textit{median}, \textit{standard deviation}, \textit{skewness}, and \textit{kurtosis} -- previous studies cited above have also used them. \added{Additionally,} to allow comparison (at least partially) \added{between point-in-time \citep[such as][]{bobra2015solar} and time series forecasting, and hence assess the potential benefits of using time series}, we consider a point-in-time feature, namely \textit{last value}. This is precisely the last value of each time series. Except for the last experiment, where fully fledged time series forecasting is attempted, all other experiments exercise point-in-time forecasting using \textit{last value}.

        Prior to computing these statistical features, the dataset requires a minimal preprocessing due to some missing values, corresponding to $<0.01\%$ of the data. Preprocessing proceeds by first filling the missing values by linear interpolation, and then using zero-one data transformation to normalize the data for our experiments. The obtained dataset of extracted features has a dimensionality of $120$, resulting from computing the $5$ above-mentioned statistics \added{(namely the four descriptive statistics of the time series and their \textit{last value})} on $24$ of the physical parameters of the SWAN-SF benchmark dataset. For several of these preprocessing steps, we employed the open-source Python package MVTS-Data Toolkit \citep{ahmadzadeh2020mvts}.
        
    \subsection{Dichotomization of Problem}\label{sec:dichotomization}
        To place the focus of our study on the challenges discussed in Section \ref{sec:theData}, we set up a simplified bi-categorical setting; we carry out experiments on binary-class data by merging X and M classes into a superclass called XM, and similarly, the classes C, B, and N into another superclass, denoted by CBN. The former becomes our minority (positive) class while the latter constitutes our majority (negative) class.

\section{Model and Treatments of Issues}\label{sec:modelAndTreatments}
    In this section, we discuss some of the challenges in training machine-learning models, caused directly or indirectly by the two most important characteristics of our dataset, namely, class-imbalance and temporal coherence. In Section \ref{subsec:controlForLearning}, we first justify why we use Support Vector Machines for analyzing the impact of different remedies for these two issues. Then, in Section \ref{subsec:svmClassifier}, for those unfamiliar with this classifier, we present a high-level discussion on how it discriminates between different classes. In Section \ref{subsec:classImbalanceTreatments} and \ref{subsec:temporalCoherenceTreatments}, we review the basic remedies for the class-imbalance and temporal-coherence issues, respectively.
    
    \subsection{Control for Learning Algorithms}\label{subsec:controlForLearning}
        Without loss of generality and for keeping our focus on the above-mentioned issues, we control for the learning algorithms by limiting our experiments to one classifier, namely, Support Vector Machines (SVM). The generality is preserved because, first, the issue of temporal coherence can only be dealt with at the data level, during preprocessing, and does not involve the classifier. The class-imbalance issue, however, can be address either at the data level or at the algorithm level, or even both. Among different data-driven strategies (e.g., sampling and feature selection) or algorithm-driven strategies (one-class learning, cost-sensitive learning, ensamble learning), only cost-sensitive learning depends on the choice of the classifier \citep{ali2013classification}. In cost-sensitive learning the class-imbalance issue is tackled by customization of the cost function so that it takes into account the imbalance ratio when penalizing misclassification. But even in this subset of remedies, although different classifiers may be impacted differently by the class-imbalance issue, they all will be impacted regardless \citep{he2009learning, krawczyk2016learning}. Therefore, the remaining of this paper can be reproduced with any other classifier without any change on the presented analyses.

    \subsection{SVM Classifier}\label{subsec:svmClassifier}
        SVMs are discriminative, supervised classifiers introduced by \citet{vapnik1963pattern}. The concept behind SVM is deeply rooted in the theory of statistical learning. The objective is to find an optimal $d$-dimensional hyperplane that separates the classes of data points, where $d$ is the number of features, determining the dimensionality of the feature space. Such a hyperplane is said to be optimal if the average distance between the plane and all the data points on the decision boundary (i.e., the support vectors) is maximum, meaning that the two classes are best segregated. For more elaborate, multi-categorical studies, a combination of multiple binary SVMs can be employed to classify the data \citep[for example, see][]{DietterichSolving1995, HastieClassification1997}. 

        The SVM classifier owes its popularity primarily to its efficient learning of nonlinear decision surfaces, thanks to its support vectors and the transformation functions (\textit{kernel}s). Different kernels can be used to provide better transformations of data into new feature spaces where the data points are potentially more accurately separable. One of the most popular kernels is \textit{Radial Basis Function} (RBF) \citep{cristianini2000an} and was found to be more effective than other kernels for this study (see Section \ref{sec:experiments}). Like any function, kernels have one or more variables that need to be specified a priori. For RBF, these are the desired smoothness of the decision surface and the radius of influence of support vectors on forming the decision surface, $C$ and $\gamma$, respectively. All kernels share the smoothness $C$ as a trade-off between performance and simplicity of the decision surface. The a priori specification of SVM hyperparameters $C$ and $\gamma$ is the required hyperparameter tuning in this case \citep[see also][for a tutorial]{burges1998tutorial}. Although this step is crucial for training an accurate and robust model and needs to be carefully carried out on every variation of the preprocessed data, in this study we only tune the hyperparameters once (on the entire dataset) and utilize their optimal values for all  experiments. This simplification is made as, again, we are not interested in finding the best models, but in highlighting the advantages and disadvantages of utilizing different preprocessing and pre-training practices.

    \subsection{Treatments of Class Imbalance}\label{subsec:classImbalanceTreatments}
        In Section \ref{sec:theData}, we discussed the class-imbalance issue and pointed to the SWAN-SF as an extremely class-imbalanced dataset. In the following, we review some of the simplest, but effective, approaches for dealing with this issue. Later in Section \ref{sec:experiments}, we verify these approaches, and using several experiments compare the impact that each of these treatments has on the robustness of the trained models.

        \begin{table*}
            \caption{Variations of undersampling (US) and oversampling (OS) approaches applied to  Partition $3$ of the SWAN-SF dataset, showing the population of classes ($\triangle$) and expansion/shrinkage factors ($\square$) in each scenario. In the header, the initial population of each class is shown ($\blacktriangle$). Methods US1, US2, and US3 are compared in Experiment F of Section \ref{subsec:impactOfSampling}. US2 and OS3 are used in Experiments A and B of Section \ref{subsec:classImbalanceImpact}, respectively.  These quantities tabulated here do not represent the number of the unique flare instances in this time period, but the number of data points collected by means of a sliding observation window with a 12-hour observation window and a 24-hour forecast window.}
            \centering
            \begin{tabular}{l | c c c c c c | l}
                \hline
                \multirow{2}{*}{\textbf{Method}}  &  & \textbf{X}    & \textbf{M}    & \textbf{C}    & \textbf{B}      & \textbf{N}     & \multirow{2}{*}{\textbf{Method Description}}\\
                                    & $\blacktriangle$ & 160  & 1152 & 3350 & 108    & 22236 & \\ \hline
                \multirow{2}{*}{US1}     & $\triangle$   & 160  & 1152 & 171  & 5      & 1135   & \multirow{2}{*}{preserves climatology in subclass level}\\ 
                                         & $\square$ & (1.00)& (1.00) & (0.05) & (0.05) & (0.05) & \\ \hline
                \multirow{2}{*}{US2}  & $\triangle$ & 160    & 160    & 106 & 106 & 106 & \multirow{2}{*}{X-based undersampling; enforces a subclass balance}\\ 
                    & $\square$ & (1.00) & (0.14) & (0.03) & (0.98) & (0.00) & \\ \hline
                \multirow{2}{*}{US3}  & $\triangle$ & 1152   & 1152   & 768 & 768 & 768 & \multirow{2}{*}{M-based undersampling; enforces a subclass balance}\\ 
                    & $\square$ & (7.20) & (1.00) & (0.23) & (7.11) & (0.03) & \\ \hline
                \multirow{2}{*}{OS1}  & $\triangle$ & 3133   & 22560  & 3350   & 108    & 22236  & \multirow{2}{*}{preserves climatology in subclass level} \\ 
                    & $\square$ & (19.58)& (19.58)&(1.00)  &(1.00)  & (1.00) &  \\ \hline
                \multirow{2}{*}{OS2}  & $\triangle$ & 1225   & 8824   & 3350   & 108    & 6592   & \multirow{2}{*}{similar to OS1 but it suppresses N} \\ 
                    & $\square$ & (7.66) & (7.66) & (1.00) & (1.00) & (0.30) & \\ \hline
                \multirow{2}{*}{OS3}  & $\triangle$ & 5025   & 5025   & 3350   & 3350   & 3350   & \multirow{2}{*}{C-based oversampling; enforces a subclass balance} \\ 
                    & $\square$ & (31.41) & (4.36) & (1.00) & (31.02)& (0.15) & \\ \hline
                \multirow{2}{*}{OS4}  & $\triangle$ & 33354  & 33354  & 22236  & 22236  & 22236  & \multirow{2}{*}{N-based oversampling; enforces a subclass balance} \\ 
                    & $\square$ & (208.46)&(28.95) & (6.64) &(205.89)& (1.00) & \\ \hline
                \hline
            \end{tabular}
            \label{tab:samplingExample}
        \end{table*}
            
        \subsubsection{Undersampling and Oversampling}
            A simple approach for tackling class imbalance in model training is to enforce class balance by either \textit{undersampling} (that is, taking out instances from the majority class) or \textit{oversampling} (that is, replicating instances of the minority class). This can achieve a nearly 1:1 balance ratio between the negative and positive samples in the training phase. This solution, however, comes at a cost: when undersampling, we leave out a significant portion of the data during training, preventing the model from taking full advantage of the entire data collection. When oversampling, on the other hand, we add replicates of the existing instances causing the model to memorize the patterns (hence overfitting), rather than generalizing. Bearing in mind the potential negative impacts caused by either of these practices, some studies recommend undersampling over oversampling \citep{drummond2003c4}. That said, one should decide about employing these models only by trial-and-error on their specific datasets. It is crucial to note that a series of more advanced, synthetic sampling techniques aiming to mitigate these impacts are long-standing \citep[e.g., see][and references therein]{chawla2002smote}. The key unknown with such methods, however, is the reliability of the generated samples, i.e., the difficulty of assessing how well the synthetic data align with the actual distributions. In our experiments, we restrict ourselves to the simpler strategies because they (1) establish lower bounds for the performance boost that can be obtained by sampling, and (2) exhibit the sampling nuances more clearly in the absence of complex assumptions that are embedded in advanced synthetic sampling strategies.
            
            While both undersampling and oversampling, in their simplest settings, seem fairly straightforward and easy to implement, one should be particularly careful when applying them to a multi-class data, such as the flare prediction dataset. This holds  despite the fact that we later (in Section \ref{sec:dichotomization}) convert our dataset to a bi-categorical class problem. The concern is that different possibilities exist depending on whether a balance at subclass level is also desired (i.e., $|\textrm{X}|=|\textrm{M}|$ and $|\textrm{C}|=|\textrm{B}|=|\textrm{N}|$). In case of undersampling, therefore, we need to decide which minority subclasses should be considered the `base' class. Letting X be the base class, we must undersample from M-class flares first to balance X and M classes and then undersample from the majority class (CBN). This yields a balanced dataset at both super-class and subclass levels. For convenience, we call this undersampling method an \textit{X-based undersampling}, since the population of X class is preserved. Similarly, other sampling methods can be introduced.  Some examples of different sampling methods on the SWAN-SF dataset are shown in Table~\ref{tab:samplingExample}.

            A quick look in  Table~\ref{tab:samplingExample} shows that the choice of the sampling method plays a critical role in tackling the classification problem. Each sampling method has to contort the flare climatology (namely, the mean occurrence frequency of flares in the time span of the dataset) in order to achieve a desired balance. This change affects the distribution of samples in the feature space by making the decision boundary either denser or sparser. A denser decision boundary makes it more difficult for the classification's hyperplane to separate the positive and negative instances correctly. Conversely, a sparser decision boundary presents an easier classification problem; i.e., with less number of misclassifications. In the bi-categorical case (XM vs. CBN), the decision boundary will most likely be where C and M classes overlap. Knowing, for example, that from the two superclasses, the C- and M-class instances are more similar to one another (than C to X, or M to B and N), it is expected that the classification problem will become somewhat easier if the chosen sampling strategy reduces the populations of either of these two classes. A good example of this situation is the oversampling method OS3 in Table \ref{tab:samplingExample}. There, X-class flares are replicated more than M-class flares (by a factor of $31.41$ vs. $4.36$) while the number of C-class flares is kept unchanged. The decision boundary, therefore, becomes sparser than it was in the unsampled data, with relatively more X- and B-class instances reducing the overall likelihood for misclassification in the training phase, hence having an easier problem in hand. Similarly for undersampling methods, both US2 and US3 yield denser decision boundaries than US1. That the sampling method US1 indeed simplifies the flare forecasting problem is verified in Experiment F of Section \ref{subsec:impactOfSampling}, and illustrated in Fig. \ref{fig:all_experiments}.
            
            Looking at Table~\ref{tab:samplingExample}, it is also important to note that in order to achieve a particular balance sometimes both undersampling and oversampling should take place simultaneously. For example, in US3, while C- and N-class instances are undersampled, X- and B-class instances are replicated roughly over 7 times, to produce the desired balance. The opposite also takes place in OS2 and OS3 methods, where N-class instances are undersampled while others are oversampled.

        \subsubsection{Misclassification Weighting}\label{subsec:misclassificationWeighting}
            Another known remedy for the class-imbalance problem is penalizing misclassification of each class differently. The SVM, like most machine learning algorithms, can incorporate different weights in its cost function. For details on how this is implemented mathematically, we refer the interested reader to \citet{ben2010user}. In this study, wherever we use misclassification weighting, we adjust the weights using the formula $w_j = \frac{n}{k \cdot n_j}$, where $n$ is the total sample size, $n_j$ is the sample size of class $j$, and $k$ represents the number of classes. To give an example, in a bi-categorical dataset with two classes, A and B, where $|A|=100$ and $|B|=900$, the weights would be $w_A=\tfrac{1000}{2 \times 100} = 5$ and $w_B=\tfrac{1000}{2 \times 900} = 0.56$. Therefore, misclassification of the minority instances (i.e., A) will be penalized $5$ times more than that in the balance case, whereas, for the majority instances, this weight actually reduces the penalty by about one half. The impact of this approach on tackling the class-imbalance issue is compared with the previously listed solutions Experiment C of Section \ref{subsec:classImbalanceImpact}.

    \subsection{Treatment of Temporal Coherence}\label{subsec:temporalCoherenceTreatments}
        Temporal coherence, as discussed in Section \ref{subsec:temporalCoherence}, appears due to the way the data was collected, and manifests itself invariably when splitting the data into the training, validation, and testing sets. To obtain a more reliable analysis on the performance and robustness of models, we often repeat this splitting process several times to verify how susceptible a model is to overfitting, and whether it has generalized well or instead memorized the patterns. This practice is called \textit{cross validation} \citep{stone1974cross, geisser1975predictive}. Cross validation is a family of statistical techniques, typically used to determine the generalization power of a model. In this process, a subset of the data, named a \textit{validation set}, is used for validation of the trained model to ensure its generalization over memorization. This subset must have no intersection with either the training set or the test set (which is reserved only for the final evaluation). There are a number of different cross-validation techniques such as $k$-fold, leave-$p$-out, stratified or purely random \citep{burman1989comparative}. Random sampling is an important part of these techniques. Disregarding the temporal coherence of the data and thus using random sampling in order to obtain the validation set may result in an artificial boost in performance and, more importantly, obscure the evidence of overfitting, i.e., the significant difference between the performance in the training and the validation phase. 
        
        To avoid this pitfall, we customize the cross validation technique such that it selects training and testing samples from different time-segmented partitions of the dataset. This, in fact, is the main reason why the SWAN-SF dataset is structured in multiple non-overlapping partitions. \added{By choosing the training and testing samples from different partitions of SWAN-SF, we prevent the model from being tested on time series which are partially identical to some of those it was trained on.} Experiment D of Section \ref{subsec:temporalCoherenceImpact}, puts this to test.
        
        In addition to the above cross-validation practice, one could use the NOAA AR numbers or HARPNUMs (i.e., unique identifiers of HARP series). This approach is already proposed in the literature \citep[e.g.,][]{bobra2015solar, florios2018forecasting, features2019Campi}, and it restricts random sampling to different tracked active regions, rather than different slices of multivariate time series possibly extracted by the same active regions. NOAA identifies and tracks active regions on the Sun's photosphere assigning NOAA AR numbers whenever possible. This is a valid treatment of temporal coherence, however, definitive HARPs often include multiple NOAA active regions or, in case of near real-time HARPs, different HARPs can be merged. In these cases, one needs to filter or untangle these complex active regions to properly tackle the temporal coherence in training, testing, and validating. Such actions are beyond the scope of the current study, therefore, we simply draw different, random samples from different partitions.

\section{Verification Measures}\label{sec:verificationMeasures}
    For a comprehensive analysis of different measures and their interpretations, we encourage the interested reader to consult \citet[][]{jolliffe2012forecast}. In this Section, however, we only briefly review the two metrics we equipped our experiments with, namely TSS and HSS2, and justify why we did so.
    
    The True Skill Statistic (TSS) \citep{hanssen1965relationship} measures the difference between the probability of detection (i.e., \textit{true-positive-rate}), $\tfrac{tp}{p}$, and the probability of false alarm (i.e., \textit{false-positive-rate}), $\tfrac{fp}{n}$, where $p = tp + fn$ and $n = fp+tn$ are the numbers of the positive and negative instances, respectively. In other words, $\textrm{TSS}=\frac{tp}{p} - \frac{fp}{n}$. It ranges from $-1$ to $+1$, where $-1$ indicates that the model is wrong in all of its predictions, $0$ means the model has no skill and reflects the success of \textit{random-guess} models (i.e., models which randomly assign labels to the instances), and $+1$ represents a perfect model that correctly assigns labels to all instances.
    
    \noindent \textbf{Advantage:} The strengths of TSS can be summarized as follows: (1) $\textrm{TSS}=0$ if and only if $tp\cdot tn=fp\cdot fn$, and this simple rule equates three seemingly different models; models that only assign positive labels, models that only assign negative labels, and random-guess models. (2) TSS includes all four elements of the confusion matrix, i.e., $tp$, $fp$, $tn$, and $fn$. (3) It is unbiased to the imbalance ratio \citep{Bloomfield_2012}. That is, given a fixed model (i.e., with an unchanging performance), regardless of the imbalance ratio, it returns a constant value representing its performance.
    
    \noindent\textbf{Disadvantage:} Using TSS has a drawback as well. Statistical models whose differences between \textit{true-positive-rate} and \textit{false-positive-rate} are the same, are not always equally good. To give a numerical example, we can calculate TSS on two models' performance: suppose we have 5,100 instances and model A gives the confusion matrix $[tp\!=\!80, fn\!=\!20, fp\!=\!0, tn\!=\!5000]$ and model B returns $[tp\!=\!90, fn\!=\!10, fp\!=\!500, tn\!=\!4500]$. While according to TSS, A and B perform equally good, i.e., $TSS(A)\!=\!TSS(B)\!=\!0.8$, model A does not misclassify any minority instance ($fp=0$), whereas in B, for every correct prediction of the minority instances ($tp$), it makes, on average, $\approx\!5.5$ incorrect predictions of that class ($fp$).
    
    \begin{figure}[t]
        \centering\includegraphics[width=\linewidth]{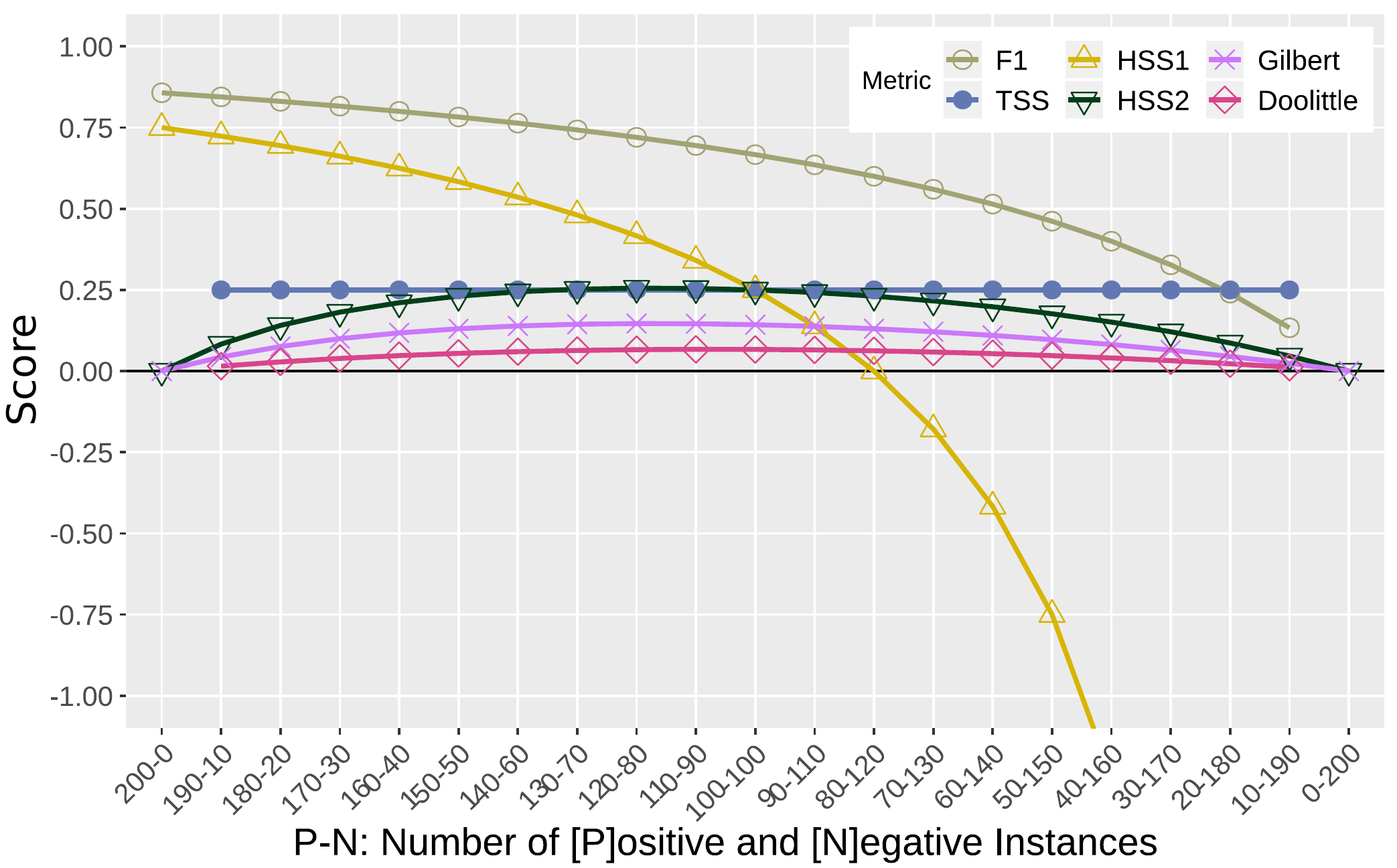}
        \caption{Comparison of several metrics' bias to the class-imbalance ratio, with the (unbiased) TSS metric in filled blue circles. For a model that correctly predicts $75\%$ of positive instances, and $25\%$ of negative instances, behavior of the performance metrics is monitored as the class imbalance transforms from $(p=0,n=200)$ to $(p=200,n=0)$, with the step-size of 10 for $p$ and $n$.}
        \label{fig:imbalanceBias}
    \end{figure}
    
    The updated Heidke Skill Score (HSS2) \citep{balch2008updated} is the other metric which quantifies the performance of a model by comparing it to the random-guess model. This is formulated as $\frac{2 ((tp \cdot tn) - (fn \cdot fp))}{p (fn + tn) + n (tp + fp)}$. Similar to TSS, HSS2 ranges within the interval $[-1, 1]$, with $0$ indicating that there is no difference between the model's performance and random-guess. Any other positive value indicates how much better than random the model of interest performs. Decreasing negative values reflect a higher similarity to a model that misclassifies all instances. The lower-bound of HSS2 approaches $-1$, as the imbalance ratio approaches 1:1.
    
    \noindent\textbf{Advantage:} HSS2 returns $0$ in all cases that TSS returns $0$. Therefore, it has the first 2 advantages of TSS while evaluating performance from a different angle. As per the numerical example given above that showed a shortcoming of TSS, HSS2 strongly favors model A over B. Precisely, $\textrm{HSS2}(A) = 0.89$, puts A much higher than B in terms of their performance, where $\textrm{HSS2}(B) = 0.23$.
    
    \noindent\textbf{Disadvantage:} Unlike TSS, HSS2 is biased to imbalance ratio.
    
    A key point in assessing metrics' appropriateness, especially in the context of rare-event classification, is their biases to the class-imbalance ratio. To highlight this, we present a comparative test designed as follows: assuming that a model correctly predicts $75\%$ of positive instances, and $25\%$ of negative instances, we monitor the behavior of a few performance metrics as we gradually change the negative-to-positive, class-imbalance ratio from 200:0 to 0:200, with the step-size of 10. In addition to TSS and HSS2, we include a few other popular metrics as well to emphasize on the unique feature of TSS. These metrics are \textit{F1-score}, HSS1,  Gilbert's Success Ratio, and Doolittle Index \citep[see their definitions in][]{jolliffe2012forecast}. Looking at the results illustrated in Fig. \ref{fig:imbalanceBias}, HSS1 and \textit{F1-score} metrics imply that the performance deteriorates as the imbalance ratio changes, which is spurious because the model's performance is assumed to remain unchanged throughout all 21 trials. Meanwhile, HSS2, Gilbert's Success Ratio, and Doolittle Index show an improvement in performance as the imbalance ratio decreases (toward the center), which is also an artifact of their susceptibility to the class-imbalance change. Among many similar metrics TSS is one of the few that remains unbiased to this changes while being informative (e.g., recall is another metric that is also unbiased but carries much less information about the performance and cannot be used as a stand-alone measure).
    
    Concluding from the above discussion, in all experiments carried out in Section \ref{sec:experiments}, we invariably provide TSS, and show HSS2 only if the class-imbalance ratio remains unchanged. These two metrics have already been reported in many flare forecasting studies with the same justification \citep[e.g.,][]{barnes2008evaluating, wilks2011statistical, bobra2015solar}. Also critically important to keep in mind is that a higher TSS value does not necessarily imply a better forecast model under the class-imbalance condition, as it may be coupled with a very low HSS2, hence a lack of robustness. In case of class balance, TSS and HSS2 (and the original HSS1) have identical values.
    
    We conclude this section by mentioning that on the topic of flare forecasting, several studies have extensively discussed the different aspects of the verification process. We avoid repeating them here but refer the interested reader to \citet[][]{barnes2008evaluating, jolliffe2012forecast, Bloomfield_2012, 2019ApJS..243...36L, cinto2020framework}.

\section{Experiments and Results}\label{sec:experiments}
    In this section, we present the experiments conducted to showcase the challenges discussed previously in the frame of the overarching flare forecasting task. We reiterate that the objective of this study is not to achieve a robust model with high performance, but to compare models trained differently. Therefore, although any new preprocessing of the data (e.g., using different normalization, data splits, or sampling techniques) require re-tuning of the hyperparameters, without loss of generality we rely on our pre-tuned hyperparameters for SVM, that is, $C=1000$, $\gamma=0.01$, with an RBF kernel. These are the results of our pre-tuning of the hyperparameters using a grid-search, by training the SVM on partitions 1, 2, and 3, and validating it on partitions 4 and 5. We compare the models' performance mainly using the TSS values achieved.

    Throughout Experiments Z, A, B, C, D, E, and F, we utilize \textit{last-value} of the time series as the single feature. In the last experiment (Experiment G), however, we use our four descriptive statistics, namely, \textit{median}, \textit{standard deviation}, \textit{skewness}, and \textit{kurtosis}, to compare their discriminative power against \textit{last-value} of the time series. This allows comparing time series with point-in-time forecasting. It is beyond the scope of this work to optimize solutions using more statistical features, so a more thorough feature selection process is deferred for future studies.

    \subsection{Baseline}\label{subsec:baseline}
        To establish a baseline for the experiments, a model first needs to learn from the available data without any special treatment in the data input process or the model configuration. With this in mind, we start with Experiment Z.

        \noindent\textbf{Experiment Z: Baseline.} This corresponds to the straightforward training of the SVM on all instances of one SWAN-SF partition and testing on another. We try this on all possible partition pairs, resulting in $20$ different trials, to expose a possible variable performance for different partition choices. The results are shown in Fig.~\ref{fig:all_experiments} (top plot; line with $\triangledown$' markers, under label `None'). The baseline TSS values across all partition pairs have a mean $\mu_{\textrm{TSS}(Z)} = 0.18$, with a standard deviation $\sigma_{\textrm{TSS}(Z)} = 0.11$.
        
    \subsection{Impact of Class-imbalance Issue}\label{subsec:classImbalanceImpact}
        Experiments A, B, and C compare the three different sampling approaches discussed in Section \ref{subsec:classImbalanceTreatments} in tackling the class-imbalance problem. The experiments share a common setup: the SVM is independently trained and tested on all permutations of partition pairs. We reiterate that each sampling method is only applied during the training phase, as modulating the sampling of the test set distorts reality and does not reflect the true operational performance of the model. To gauge the confidence of a model's performance when a sampling method is employed, we repeat the experiment 10 times and report the mean and standard deviation values of the achieved TSS. 
        
\begin{figure*}[hbtp]
        \centering
        \includegraphics[width=0.93\textwidth]{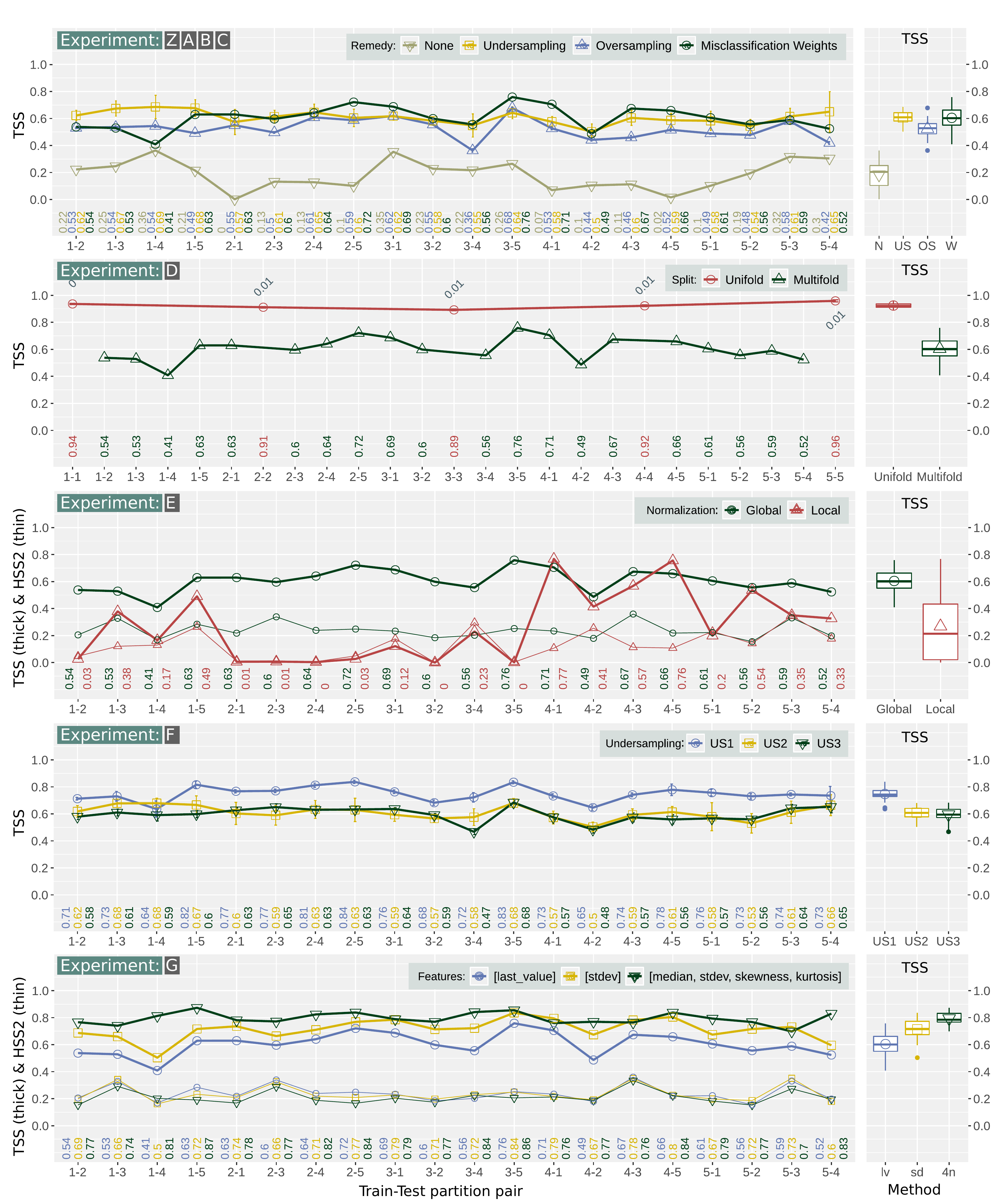}
            \caption{Performance of SVM (by means of TSS/HSS2), trained and tested on partition pairs of the SWAN-SF benchmark dataset. Error bars show the standard deviation of individual TSS/HSS2 values, produced after 10 times of repeating each experiment. The TSS value corresponding to each partition pair is written above the horizontal axis. On the right side of each plot, the box plots depict the overall variation (interquartile range (IQR), median, and mean) of TSS across all pairs, for each scenario. The experiment(s) each plot corresponds to are mentioned on the top-left corner of each plot. In Experiments E and G, thinner lines correspond to HSS2, in addition to TSS.}
            \label{fig:all_experiments}
    \end{figure*}

        \noindent\textbf{Experiment A: Undersampling.} In the training phase, the model takes in a subset of a training partition generated by an X-based undersampling method (US2 from Table~\ref{tab:samplingExample}). This enforces a 1:1 balance not only in the super-class level (i.e., $|\textrm{XM}|=|\textrm{CBN}|$) but also in the subclass level (i.e., $|\textrm{X}|=|\textrm{M}|$ and $|\textrm{C}|=|\textrm{B}|=|\textrm{N}|$). The trained model is then tested against all other partitions. In the top plot in Fig.~\ref{fig:all_experiments} (line with `$\square$' markers, under label `Undersampling'), the consistent and significant impact of this remedy is evident, when compared to the baseline. We obtain a mean TSS  $\mu_{\textrm{TSS}(A)} = 0.61$ with standard deviation $\sigma_{\textrm{TSS}(A)} = 0.05$, to be compared with 0.18 and 0.11, respectively, of the baseline Experiment Z. 

        \noindent\textbf{Experiment B: Oversampling.} Here we apply a C-based oversampling method (OS3 in Table~\ref{tab:samplingExample}) that also enforces a 1:1 balance in the subclass level. As shown in the top plot in Fig.~\ref{fig:all_experiments} (line with `$\triangle$' markers, under label `Oversampling'), we obtain a mean TSS $\mu_{\textrm{TSS}(B)} = 0.52$ with standard deviation $\sigma_{\textrm{TSS}(B)} = 0.07$. Notice that undersampling and oversampling give similar TSS values, with a difference within applicable uncertainties.

        \noindent\textbf{Experiment C: Misclassification Weighting.} We use the imbalance ratio of the superclasses as the weights (as discussed in Section \ref{subsec:misclassificationWeighting}). For instance, when working with Partition 3, since the majority-to-minority ratio is 20:1, we set the misclassification weights as follows: $w_{\textrm{XM}}=20$ and $w_{\textrm{CBN}}=1$. The top plot in Fig.~\ref{fig:all_experiments} (line with `$\circ$' markers, under label `Misclassification Weights') shows that misclassification weighting typically scores near the top of the achieved TSS values ($\mu_{\textrm{TSS}(C)} = 0.60$ and $\sigma_{\textrm{TSS}(C)} = 0.09$), albeit within uncertainties from undersampling ($\mu_{\textrm{TSS}(A)} = 0.61$) and oversampling ($\mu_{\textrm{TSS}(B)} = 0.52$) remedies. This said, it is worth pointing out that this remedy is probably better suited toward producing more robust forecast models because it allows a data-driven tunability of applicable weights.
    
    \subsection{Impact of Temporal Coherence}\label{subsec:temporalCoherenceImpact}
        In Section \ref{subsec:temporalCoherence}, we discussed the theoretical impact of random sampling, embedded in many cross validation methods, on a temporally coherent dataset. The following experiment is designed to quantify this impact.

        \noindent\textbf{Experiment D: Data Splits.} This time, the SVM is trained and tested on two randomly chosen subsets of the same partition, with no overlap between the subsets, while preserving the climatology of the flares in each of them. More accurately, this is a stratified, $k$-fold cross validation using a random sub-sampling method with $k=10$. The results are juxtaposed with those obtained by training SVM on one partition and testing it on another. In both scenarios, we equipped SVM with misclassification weights, to eliminate the need for an additional sampling layer. Therefore, the only determining factor is whether the instances are sampled from the same partition or not.

        The second plot in Fig.~\ref{fig:all_experiments} illustrates this comparison. When SVM is trained and tested on a single partition (line with `$\circ$' markers, under label `unifold'), performance is boosted significantly with TSS averaging at $\mu_{\textrm{TSS}(D_u)}=0.92$, with a standard deviation of $\sigma_{\textrm{TSS}(D_u)}=0.03$. Training and testing on different partitions (line with `$\triangle$' markers, under label `multifold') shows $\mu_{\textrm{TSS}(D_m)}=0.60$, with a standard deviation  $\sigma_{\textrm{TSS}(D_m)}=0.09$. Subscripts \textit{u} and \textit{m} here correspond to unifold and multifold training/testing, respectively. The remarkable difference between the two scenarios is because the forecast models in unifold trials are both trained and tested on a temporally coherent dataset. Although for the unifold scenario the training and testing sets are non-overlapping, random sampling does not properly treat the temporal coherence and consequently, the models show such a high performance. Such results are occasionally misinterpreted as evidence of models' robustness because of the consistency between the training and testing results. However, as discussed in Sections \ref{subsec:temporalCoherence} and \ref{subsec:temporalCoherenceTreatments}, they are indications of overfitting due to memorization.

    \subsection{Impact of Normalization: Global and Local}\label{subsec:normalizationImpact}
        \noindent\textbf{Experiment E: Normalization.} As we train and test SVM on a pair of partitions, we transform the feature space to a normalized space using a zero--one normalization method. This can be done either by means of features' extrema of both partitions, or by taking into account the extrema of each partition separately. We refer to the former as \textit{global} normalization, and to the latter as \textit{local} normalization. It is worth noting that data normalization is an important preprocessing step in data cleaning, that needs a rigorous analysis of the data, including finding the outliers and obtaining the statistically meaningful extrema for each of the features. The current experiment only aims to compare the performance of the two normalization approaches for different partitions, hence phases of the solar cycle.
        
        A comparison of different performances, shown in the third plot in Fig.~\ref{fig:all_experiments}, reflects the significant changes in performance due to the continuously modulating magnetic activity during the 11-year solar cycle. Mean TSS values of the global (line with `$\circ$', under label `Global') and local (line with `$\triangle$', under label `Local') normalizations 
        are $\mu_{\textrm{TSS}(E_g)}=0.60$ and $\mu_{\textrm{TSS}(E_l)}=0.27$, respectively. The relatively high standard deviation of 
        $\sigma_{\textrm{TSS}(E_l)}=0.26$ in case of local normalization, compared to $\sigma_{\textrm{TSS}(E_g)}=0.09$ for the global normalization implies that the local normalization is generally inconsistent between the training and the testing partitions. Statistically, higher performance is achieved when global normalization is applied, although in some cases the opposite happens, according to TSS, HSS2, or both. Also, $\mu_{\textrm{HSS2}(E_l)} = 0.13$, $\mu_{\textrm{HSS2}(E_g)} = 0.24$, and $\sigma_{\textrm{HSS2}(E_l)} = 0.11$, $\sigma_{\textrm{HSS2}(E_g)} = 0.06$. 

        It is interesting to discuss the model’s performance for some specific partition pairs. For the local normalization instances, it seems that Partition 4 (that spans over the period of March 2014 through March 2015) lays out a unique set of extrema with representative features for all partitions, that results in achieving a slightly better performance than global normalization. In contrast, Partition 2 (February 2012 through October 2013) seems to be holding the most restrictive range of values. A potential justification for this difference could be the presence of more extreme outliers in Partition 2, that corresponds to solar maximum compared to Partition 4, that corresponds to the decay phase of solar cycle 24. This suggests that a careful outlier detection process in the data cleaning phase could also improve, or otherwise affect, the overall performance of forecast models.

        While additional investigation of this effect is  clearly warranted, one gathers that performance may be optimized when training and testing take place during similar levels of solar activity, when local and global normalization converge on similar normalized values. This, however, is hardly tenable in operational settings, when forecast models perform on an always unknown day-by-day solar activity level.

    \subsection{Impact of Sampling}\label{subsec:impactOfSampling}
        In Section \ref{subsec:classImbalanceTreatments}, we showed that there are multiple variants of oversampling and undersampling strategies, a subset of which are listed in Table.~\ref{tab:samplingExample}. We also presented how this affects flare distributions in each partition of the SWAN-SF benchmark. Below, we test how different undersampling methods impact models' performance in terms of their TSS values, across different partitions. Testing different oversampling methods would essentially teach us similar lessons.

        \noindent\textbf{Experiment F: Different Undersampling Methods.} We use all undersampling methods listed in Table~\ref{tab:samplingExample}, namely, US1, US2, and US3, in the training phase. Then we test the trained models against all other partitions. All three undersampling methods yield a 1:1 balance in the super-class level (i.e., $|XM| = |CBN|$). However, US1 preserves the climatology of flares in the subclass level, while US2 and US3 additionally enforce a 1:1 balance in the subclass level, as well. US2 and US3 reduce the flare populations based on the number of X-class and M-class flares, respectively. We further employ global normalization only to keep the experiments' conditions constant. Our results are shown in Fig.~\ref{fig:all_experiments}, the fourth plot from top. From them, one sees a relatively similar, consistent performance, although the climatology-preserving undersampling, i.e., US1 (line with `$\circ$', under label `US1') seems to give a statistically higher performance compared to the other two methods. This is largely because in both US2 and US3 the number of flare-quiet instances are substantially reduced to meet the required criterion of 1:1 balance in the subclass level. As discussed in Section \ref{subsec:classImbalanceTreatments}, this takes away many easy-to-predict instances from the models, and makes the problem more challenging, hence the lower performance for US2 and US3.
        
        Concluding form this experiment, it becomes clear that different undersampling and oversampling methods give non-identical performances, and if one prefers to use undersampling or oversampling over misclassification weighting strategy, we recommend the climatology-preserving class-balancing for better robustness on the SWAN-SF (and, apparently, similar benchmark datasets.)

    \subsection{Impact of Time Series Features}\label{subsec:otherFeatures}
        We reserve the last experiment for presenting the benefit of using time series, rather than point-in-time values, for forecasting. To simulate the point-in-time effect we employ the \textit{last value} statistic that returns the last record of each time series of the SWAN-SF, converting a multivariate time series to a vector of \textit{last value}s.

        \noindent\textbf{Experiment G: SVM With Other Statistical Features.} SVM is trained and tested on different partition pairs, using three sets of \added{statistics extracted from the time series. These statistics sets are:} (i) \{\textit{last value}\}, (ii) \{\textit{standard deviation}\}, and (iii) \{\textit{median}, \textit{standard deviation}, \textit{skewness}, \textit{kurtosis}\}. In all these three cases, misclassification weighting is used as a class-imbalance remedy. As the box plot in the bottom plot in Fig.~\ref{fig:all_experiments} shows, \textit{standard deviation} (line with `$\square$' markers, under label `[stdev]') results in a statistically better performance than \textit{last value} (line with `$\circ$' markers, under label `[last\_value]'), while the third set (line with `$\triangledown$' markers, under label `[median, stdev, skewness, kurtosis]') seems to outperform \textit{standard deviation}. This is a good indication that different characteristics of time series carry important, non-redundant information that may improve the reliability of a forecast model.

        This said, note that HSS2 does not subscribe to such differences. This is a good example illustrating why the choice of metric must be specific to the objective of the task. Our interpretation of the apparent improvement here relies solely on how TSS reflects success. HSS2, on the other hand, does not show any significant improvement among the three choices. Therefore, if the objective of a task is more aligned with what HSS2 measures, then the three approaches taken in this experiment are very similar, within the margin of error. To improve HSS2 as well as TSS, more time needs to be invested on tuning the SVM's hyperparameters. For the reasons mentioned at the beginning of Section \ref{sec:experiments} we did not tune SVM separately for each experiment. Also note that the utilized statistics are only chosen due to their general descriptive power and are not the outcome of a rigorous feature-selection process. Finding a set of statistical features that optimally distinguish between metadata time series of flaring and non-flaring active regions is in itself a challenging task and goes beyond the scope of this work.

\section{Summary, Conclusions, and Future Work}\label{sec:summary}
    Working with multi-class, high dimensional data is certainly challenging. From time to time and despite several revisions of these challenges, they tend to be overlooked by domain experts, understandably so, as the complexity of the original flare forecasting problem may overshadow the challenges in the preprocessing of the data. We used the SWAN-SF benchmark dataset as a reference in order to highlight some of the challenges. We pointed attention to an interesting characteristic of such datasets, that we called \textit{temporal coherence}, inherited from the sampling and slicing methodologies. We also revisited the problem of class imbalance in flare forecasting and the impact of different remedies addressing this problem. In the context of temporally coherent, class-imbalanced data, we designed several different, non-overlapping experiments showcasing these challenges and common preprocessing tasks to tackle them, in terms of normalization, sampling, and cross-validation. Below, we summarize our key points and conclusions:

    \begin{enumerate}
        \item \textbf{On Normalization:} A global normalization of parameters (i.e., over both training and testing partitions) is preferred over a local normalization (i.e., separately over training and testing partitions). However, since the variance of the physical parameters changes significantly from one partition to another, following the changes within a solar cycle, a universal normalization (i.e., over the entire five partitions) should be avoided. In other words, the choice of the extrema should take into account the phase of the solar cycle in order to achieve the optimal forecasting capability. This said, we realize that the unavoidable local normalization in operational settings may have an impact on performance. In addition, our approach ignores the actual values of metadata parameters, some of which have also shown flare predictive ability in numerous previous studies cited in the Introduction.
        
        \item \textbf{On Class Imbalance:} The class-imbalance problem needs to be dealt with properly, i.e., using simple or synthetic sampling approaches, weighting, or other means. Regarding sampling, any methodology applies only to the training set: altering the climatology of flares in the test and validation sets leads to unrealistic and overly optimistic performance that cannot be reproduced in operational settings. We recommend treating the class-imbalance problem by means of misclassification weighting, if the cost function of the utilized model allows.
        
        \item \textbf{On Temporal Coherence:} Any random sampling employed for splitting data into training, validation, and test sets must take into account the temporal coherence of data that comes into play when the slicing time step is smaller than the observation window in time series forecasting. This is on top of fulfilling the non-overlapping condition dictating that a given instance must appear in one and only one of the three sets. The time-segmented SWAN-SF dataset with its non-overlapping partitions can help models circumvent these issues. Another approach is to limit the random sampling to unique HARP series or NOAA active regions, as studied in detail by \citet{features2019Campi}, among others. 
        
        \item \textbf{On Performance Metrics:} In solar flare forecasting, class imbalance is intrinsic and at least some of the metrics chosen for verification of performance should not be susceptible to or biased by it. As previous studies have shown \citep[such as, for example][]{woodcock1976evaluation,Bloomfield_2012,bobra2015solar} TSS is not affected adversely by class imbalance. Hence, we also recommend using this metric for flare prediction, at least in binary (i.e., non-probabilistic) forecasting. This said, HSS2 should also be used as it quantifies how a model performs in comparison to the random-guess model. HSS2 is affected by class imbalance, however, in perfectly balanced data $\textrm{HSS1}=\textrm{HSS2}=\textrm{TSS}$. It is also possible that a specific task (with a particular objective) may call for a different, or even a new, metric.
        
        \item \textbf{On Comparison of Models:} Different models (i.e., fitted algorithms) are not precisely comparable unless they (1) are  trained on the same dataset, (2) with identical normalization techniques, (3) with identical sampling strategies, and (4) with identical class-imbalance remedies, if any. The use of a benchmark dataset, allows satisfying the first condition towards producing comparable studies on flare forecasting. In this study, we highlighted the importance of the other three conditions.
    \end{enumerate}
    
    There are many further interesting avenues to be discovered that we plan to exploit in future studies. Results reported here are based on a small set of five time series statistics (\textit{median}, \textit{variance}, \textit{skewness}, \textit{kurtosis}, and \textit{last-value}). There was no consistent effort to find the best-performing features that could potentially boost the models' performance. Some of them could be based on, for example, the first and second derivatives of the time series, those that keep track of the general trends in time series, up-surges and down-slides, features that compare different parts of the time series, and many others. To be able to benefit from a long list of features, a feature selection algorithm is needed that must be carefully guided and augmented by domain expertise. Time series of the best selected features could, besides improving flare forecasting in operational settings, help solar physicists better understand the triggering mechanism(s) of solar flares. Interpretable machine learning methods, rather than `black-box' solutions, could be paramount for this purpose and should be given priority \citep{lipton2018troubling, marcus2018deep,rudin_stop_2019}. Hard-to-interpret methodologies often based on deep learning may, on the other hand, be problematic for solar flare forecasting, given the limited amount of data available, where millions of positive sample instances are typically needed \citep[e.g.,][]{Goodfellow-et-al-2016}.

    In such efforts, well-curated benchmark datasets have been found to be instrumental. With the SWAN-SF, for example, we have and will continue to run large numbers of experiments in relatively short time and with considerable efficiency. Benchmark datasets, adhering to well-defined data collection and integration policies, can always be further enhanced and expanded, offering valuable services to interested communities. We hope that this and other recent works raise awareness within this and other domains dealing with class imbalance and temporal coherence in their respective forecasting problems \citep{2019AGUFMSH34B..05S, 2019AGUFMSH34A..01H, 2019AGUFMNG22A..06A, 2020arXiv200612224N}. Interdisciplinarity is a key common element of such pursuits, so extended applicability may provide clues, or even viable solutions, for future diverse forecast efforts and other real-world problems.

\section*{Acknowledgments}
    We would like to thank the anonymous reviewer for the helpful feedback they provided on earlier drafts of the manuscript.
    
    This work was supported in part by two NASA Grant Awards [No. NNH14ZDA001N, 80NSSC20K1352], and two NSF Grant Awards [No. AC1443061 and AC1931555]. The AC1443061 award has been supported by funding from the Division of Advanced Cyber infrastructure within the Directorate for Computer and Information Science and Engineering, the Division of Astronomical Sciences within the Directorate for Mathematical and Physical Sciences, and the Division of Atmospheric and Geospace Sciences within the Directorate for Geosciences.


\pagebreak

\appendix

    For convenience, all acronyms and notations used in this work are listed in Tables \ref{tab:acronyms} and \ref{tab:notations}.
    
    \begin{table}[htp]
        \caption{A list of all Acronyms used in this manuscript.}
        \centering\footnotesize
        \begin{tabular}{ p{2cm} | p{12cm} }
            \hline
            \textbf{Acronym}      &  \textbf{Full Name}\\\hline
            SDO          &  National Oceanic and Atmospheric Administration\\\hline
            GOES         &  Geostationary Operational Environmental Satellite\\\hline
            SWAN-SF      &  Space Weather data ANalytics for Solar Flare; a benchmark data produced by Georgia State University's Astroinformatics Cluster\\\hline
            SVM          &  Support Vector Machine; a family of statistical learning models\\\hline
            TSS          &  True Skill Score \citep{hanssen1965relationship}\\\hline
            HSS1         &  Heidke Skill Score \citep{heidke1926berechnung}\\\hline
            HSS2         &  Updated Heidke Skill Score \citep{balch2008updated}\\\hline
            GS           &  Gilbert's Success Ratio, also known as Gilbert Skill Score or Equitable Threat Score \citep{gilbert1884finley}\\\hline
            DI           &  Doolittle Index \citet{doolittle1885verification}\\\hline
            US$_i$       &  An undersampling method indexed $i$-th in Table~\ref{tab:samplingExample}\\\hline
            OS$_i$       &  An oversampling method indexed $i$-th in Table~\ref{tab:samplingExample}\\\hline
            \hline
        \end{tabular}
        \label{tab:acronyms}
    \end{table}
    
    \begin{table*}[htp]
        \caption{A list of all notations used in this manuscript.}
        \centering\footnotesize
        \begin{tabular}{p{2cm} | p{12cm} }
            \hline
            \textbf{Notation}     &  \textbf{Description}\\\hline
            X, M, C, B, A   & Used both as the GOES flare class label and as the set of all instances of one class.\\\hline
            N                       & Used both as the non-flaring class label and as the set of all non-flaring instances.\\\hline
            XM                      & Used both as the flaring class label (instances from X and M) and as the set of all X- and M-class instances.\\\hline
            CBN                     & Used both as the non-flaring class (instances from C, B, and N) and as the set of all C-, B-, and N-class instances.\\\hline
            $|\cdot|$                 & Set cardinality; e.g., $|X|$ indicates the number of X-class flares.\\\hline
            $T_{obs}$                 & Temporal observation window; a time interval during which the values of a set of magnetic-field parameters are observed.\\\hline
            $T_{pred}$                & Temporal prediction window; a time interval that is labeled according to the strongest flare class recorded in that period.\\\hline
            $\tau$                    & Temporal step size in the moving-window approach used for slicing the time series.\\\hline
            $t_i$                     & Time series with index $i$.\\\hline
            $e$                      & Statistical event, such as \textit{flaring} or \textit{non-flaring}.\\\hline
            $\mathcal{P}(e)$         & Probability of the statistical event $e$.\\\hline
            $p$                      & Positive class label of an instance; in flare-forecasting context, it points to a data point that is predicted as flaring.\\\hline
            $n$                      & Negative class label of an instance; in flare-forecasting context, it points to a data point that is predicted as non-flaring.\\\hline
            $tp$                     & True positive (\textit{hit}); a correct classification of an instance that is labeled as positive.\\\hline
            $fp$                     & False positive (\textit{false alarm}); an incorrect classification of an instance that is labeled as negative.\\\hline
            $tn$                     & True negative (\textit{correct rejection}); a correct classification of an instance that is labeled as negative.\\\hline
            $fn$                     & False negative (\textit{miss}); an incorrect classification of an instance that is labeled as positive.\\\hline
            \hline
        \end{tabular}
        \label{tab:notations}
    \end{table*}


\bibliography{howToTrain.bib}{}
\bibliographystyle{aasjournal}

\listofchanges

\end{document}